\documentclass[aip,cha, reprint,showpacs,showkeys]{revtex4-1}
\usepackage{amssymb, dcolumn,  graphicx, latexsym, amsfonts, bm}

\begin{document}

\title{The parametric instability in the inductively coupled plasma driven by the ponderomotive current.}
\author{V. V. Mikhailenko}\email[E-mail: ] { vladimir@pusan.ac.kr}
\affiliation{Plasma Research Center, Pusan National University, Busan 46241, South Korea.}
\author{V. S. Mikhailenko}\email[E-mail:] {vsmikhailenko@pusan.ac.kr}
\affiliation{Plasma Research Center, Pusan National University, Busan 46241, South Korea.}
\author{H. J. Lee}\email[E-mail:] {haejune@pusan.ac.kr}%
\affiliation{Department of Electrical Engineering, Pusan National University, Busan 46241, 
South Korea.}

\begin{abstract}
The stability theory of the skin layer plasma of the inductive discharge is developed for the case when the electron 
quiver velocity in RF wave is of the order of or is larger than the electron thermal velocity. This theory is grounded 
on the methodology of the oscillating modes, which accounts for the oscillation motion of the electron component relative 
to the unmovable ions in the spatially inhomogeneous RF field of the skin layer. The theory predicts the existence 
the instability of the parametric type in a skin layer with the growth rate comparable with frequency. This instability stems 
from the coupled action of two effects caused by the electron-ion relative motion in RF field: occurrence of harmonics 
of the perturbed potential and their coupling due to the ponderomotive current. The instability exists in the finite 
interval of the ponderomotive current velocity and is absent in the uniform boundless plasma.
\end{abstract}

%\noindent{\it Keywords\/ kinetic theory, drift Alfven instability, shear flow}

\maketitle

\section{Introduction}\label{sec1}
The regime of the anomalous skin effect (or nonlocal regime)\cite{Weibel, Kolobov} is typical for the low pressure inductive plasma sources 
employed in material processes applications\cite{Lieberman}. It occurs when the frequency $\omega_{0}$ of the operating 
electromagnetic (EM) wave is much above the electron-neutral collision frequency, but less than the electron plasma frequency. 
In this regime, the interaction of the EM field with electrons is governed by the electron thermal motion. For this reason, the EM wave absorption
\cite{Shaing}, the formation of the anomalous skin layer near the plasma boundary\cite{Weibel}, and the anomalous 
electron heating\cite{Aliev, Tyshetskiy} require the kinetic description which involves the well known mechanism of collisionless power dissipation - Landau 
damping. It stems from the resonant wave-electron interaction under condition that the electron thermal velocity $v_{Te}$ is comparable with (or is larger 
) the EM phase velocity. The theory of the anomalous skin effect is developed as a rule, employing the linear approximation to the solution of the 
Vlasov equation for the electron distribution function. It is assumed in this theory that the equilibrium electron distribution function depends only 
on electron kinetic energy and does not involve the electron motion in the time dependent spatially inhomogeneous EM wave. 
This approximation is valid when the quiver velocity of electron in the EM wave is negligible in comparison with the electron thermal velocity.

It was found experimentally \cite{Godyak1, Godyak2, Godyak3, Godyak4} and analytically\cite{Cohen1,Cohen2,Smolyakov1,Piejak,Smolyakov2} 
that at the low driving frequency of an inductive discharge, at which RF Lorentz force acting on electrons becomes comparable 
to or larger than the RF electric field force, the nonlinear effects in the skin layer becomes essential. The theoretical analysis 
\cite{Cohen1,Cohen2, Smolyakov1,Piejak,Smolyakov2}  has shown that electron oscillatory motion in the inhomogeneous RF field in the skin layer leads to 
the ponderomotive force. This force is regarded as the responsible for the reduction of the steady state electron density distribution within the 
skin layer\cite{Cohen1,Cohen2} and  for the formation of experimentally observed\cite{Godyak1, Godyak2, Godyak3, Godyak4} and analytically predicted
\cite{Smolyakov1,Piejak,Smolyakov2} second harmonics which was found \cite{Godyak1, Godyak2, Godyak3, Godyak4} to be much larger than the electric field on 
the fundamental frequency.

Our paper is devoted to the analytical investigations of the nonlinear processes  in the skin layer in the case
of high frequency of the operating EM wave for which the RF electric field force acting on electrons prevails over the RF Lorentz
force. In this case, a situation can occur that the electron quiver velocity in a skin layer under the action of the electromagnetic wave 
approaches or is larger than the electron thermal velocity. 
Under such conditions it is reasonable to talk about free oscillations of a plasma particle 
under the action of the RF field (at least, in the zero approximation in the ratio of the 
collision frequency to the field frequency). 
The relative oscillatory motion of the electrons and ions in the RF field is a potential 
source of numerous instabilities 
of the parametric type (see, for example, Refs.\cite{Silin,Porkolab,Mikhailenko1,Akhiezer}) with frequencies $\omega$  comparable with or 
less than the frequency $\omega_{0}$ of the applied RF wave. It is clear that in such a situation  an essentially 
nonlinear dependence of the plasma conductivity on the RF field as well as the anomalous absorption of the RF energy 
due to the development of the plasma turbulence and turbulent scattering of electrons arise. 
This is just the case which interest us in the present paper.

It is usually accepted in the theoretical investigations of the parametric instabilities 
excited by the strong RF wave in the unbounded uniform plasmas, 
that the approximation of the spatially homogeneous pump wave may suffice since the parametrically 
excited waves have the wave number much larger than the wave number of the primary RF wave. The 
presence of the skin layer at the plasma boundary where RF wave decays into plasma requires the development of new 
approach to the theory of the instabilities of the parametric 
type in  which the spatial inhomogeneity of the RF wave should be accounted for. This new 
kinetic approach, grounded on the methodology of the oscillating 
modes, is developed in this paper and is presented in Sec. \ref{sec2}. We found that electrons in the skin layer experience 
the oscillating motion in EM field jointly with the uniformly accelerated motion under the action of the 
ponderomotive force which stems from the spatial inhomogeneity of the RF wave EM field. The basic equation for the perturbed 
electrostatic potential which determines the stability of the inductively 
coupled plasma against the development of the electrostatic instabilities in the skin layer is 
derived in Sec. \ref{sec3}. The numerical solution of this 
equation is presented in Sec. \ref{sec4}. It reveals the instability which is the result of 
the coupled action of the oscillating and steady 
motion of the electrons relative to the ions. Conclusions are presented in Sec. \ref{sec5}. 

\section{Basic transformations and governing equations}\label{sec2}
We consider a model of a plasma occupying region $z\geqslant 0$. The RF antenna which 
launches the RF wave with frequency $\omega_{0}$ is assumed to exist to the left of the plasma boundary $z=0$. The electric, $
\mathbf{E}_{0}\left(z,t\right)$, and magnetic, $\mathbf{B}_{0}\left(z,t\right)$, fields of a 
such RF wave, are directed along the plasma boundary and 
attenuate along $z$ due to the skin effect. We assume that these fields are exponentially 
decaying with $z$, and sinusoidally varying with time, 
\begin{eqnarray}
&\displaystyle 
\mathbf{E}_{0}\left(z,t\right)=E_{0y}e^{-\kappa z}\sin \omega_{0}t\mathbf{e}_{y},
\label{1}
\end{eqnarray}
and 
\begin{eqnarray}
&\displaystyle 
\mathbf{B}_{0}\left(z,t\right)=E_{0y}\frac{c\kappa}{\omega_{0}}e^{-\kappa z}\cos \omega_{0}t\mathbf{e}_{x},
\label{2}
\end{eqnarray}
where $\mathbf{E}_{0}$ and $\mathbf{B}_{0}$ satisfy the Faraday’s law, $\partial E_{0}/
\partial z=\partial B_{0}/c\partial t$. In this paper, we consider the effect of the relative motion of plasma species in the applied RF field on the 
development the short scale electrostatic perturbations in the skin layer  with wavelength much less than the skin layer thickness. 
Our theory bases on the Vlasov equation for the velocity distribution function $F_{\alpha}$ of $\alpha$ species 
($\alpha=e$ for electrons and $\alpha=i$ for ions), 
\begin{eqnarray}
&\displaystyle 
\frac{\partial F_{\alpha}}{\partial t}+\mathbf{v}\frac{\partial F_{\alpha}}
{\partial\mathbf{r}}+\frac{e_{\alpha}}{m_{\alpha}}\left(\mathbf{E}_{0}
\left(z,t\right)+\frac{1}{c}\left[\mathbf{v}\times\mathbf{B}_{0}\left(z,t \right)\right] \right.
\nonumber
\\ 
&\displaystyle
-\nabla \varphi\left(\mathbf{r},t\right) 
\left)\frac{\partial F_{\alpha}}{\partial\mathbf{v}}\right. =0.
\label{3}
\end{eqnarray}
This equation contains the potential $\varphi\left(\mathbf{r},t\right)$ of the electrostatic plasma perturbations which is 
determined by the Poisson equation
\begin{eqnarray}
&\displaystyle \vartriangle \varphi\left(\mathbf{r},t\right)=
-4\pi\sum_{\alpha=i,e} e_{\alpha}\int f_{\alpha}\left(\mathbf{v},
\mathbf{r}, t \right)d\textbf {v}_{\alpha}, \label{4}
\end{eqnarray}
where  $f_{\alpha}$ is the perturbation of the equilibrium distribution function $F_{0\alpha}$,
$F_{\alpha}=F_{0\alpha}+f_{\alpha}$. $F_{0\alpha}$ is a function of the canonic momentums $p_{z}=m_{\alpha}v_{z}$ and 
$p_{y}=m_{\alpha}v_{y}-\frac{e}{c}A_{0y}\left(z, t\right)$, which are the integrals of the Vlasov equation (\ref{3}) without potential $\varphi
\left(\mathbf{r},t\right)$. It will be assumed to have a form
\begin{eqnarray}
&\displaystyle 
F_{\alpha 0}\left(v_{y},v_{z},z,t \right) =\frac{n_{0\alpha}}{2\pi v^{2}_{T\alpha} }
\exp\left[-\frac{v^{2}_{z}}{2v^{2}_{T\alpha}}\right.
\nonumber
\\ 
&\displaystyle
\left.-\frac{1}{2v^{2}_{T\alpha}}\left(v_{y}
-\frac{e}{cm_{\alpha}}A_{0y}\left(z,t \right) \right)^{2} \right],
\label{5}
\end{eqnarray}
where the electromagnetic potential $A_{0y}\left(z,t \right)$ for the EM field (\ref{1}) and (\ref{2}) is equal to\cite{Cohen1, Cohen2} 
\begin{eqnarray}
&\displaystyle 
\mathbf{A}_{0y}\left(z,t \right)= \frac{c\mathbf{E}_{0y}}{\omega_{0}}e^{-\kappa z}\cos \omega_{0}t.
\label{6}
\end{eqnarray}.

The kinetic theory of the plasma stability with the time dependence of the $F_{\alpha 0}$ caused by 
the strong spatially homogeneous oscillating RF electric field $\mathbf{E}_{0}\left(t\right)=E_{0y}
\sin \omega_{0}t\mathbf{e}_{y}$ was developed\cite{Silin, Porkolab} by employing the transformation $\mathbf{v}=\mathbf{v}_{\alpha}
+\mathbf{V}_{\alpha 0}\left(t\right)$ of the velocity  $\mathbf{v}$ in the Vlasov equations for ions and electrons to velocity 
$\mathbf{v}_{\alpha}$ determined in the frame of references which oscillates with velocity $\mathbf{V}_{\alpha 0}\left(t\right)$ of particles of species 
$\alpha$  in velocity space, leaving unchanged the position coordinates. With new velocity $\mathbf{v}_{\alpha}$ the explicit time dependence which 
stems from the RF field is excluded from the Vlasov equation. In this paper we employ more general transformation of the velocity and position 
coordinates to the convected-oscillating frame of references determined by the relations
\begin{eqnarray}
&\displaystyle 
\mathbf{v}_{\alpha}=\mathbf{v}-\mathbf{V}_{\alpha}\left(\mathbf{r},t \right) , 
\nonumber
\\ 
&\displaystyle
\mathbf{r}_{\alpha}=\mathbf{r}-\mathbf{R}_{\alpha}\left(\mathbf{r},t \right)
= \mathbf{r} -\int\limits^{t} \mathbf{V}_{\alpha}\left(\mathbf{r},t_{1} \right)dt_{1}.
\label{7}
\end{eqnarray}
This transformation was decisive in the development of the parametric weak turbulence theory
\cite{Mikhailenko1}, and the theory of the stability and turbulence of plasma in RF wave with finite wavelength\cite{Mikhailenko2} and admits the 
solution of the Vlasov equation in the case of the oscillating spatially inhomogeneous RF field. The transformation of Eq. (\ref{3}) for $F_{e}$
to velocity $\mathbf{v}_{e}$ and coordinate $\mathbf{r}_{e}$ variables determined by Eq. (\ref{7}) transforms Eq. (\ref{3}) to the form
\begin{eqnarray}
&\displaystyle 
\frac{\partial F_{e}\left(\mathbf{v}_{e},\mathbf{r}_{e},t \right) }{\partial t}
+\mathbf{v}_{e}\frac{\partial F_{e}}{\partial\mathbf{r}_{e}}
-v_{ej}\int\limits^{t}_{t_{0}}\frac{\partial V_{ek}\left( \mathbf{r},t_{1}\right) }{\partial r_{j}}dt_{1}
\frac{\partial F_{e}}{\partial r_{ek}}
\nonumber
\\ 
&\displaystyle
-v_{ej}\frac{\partial V_{ek}}{\partial r_{j}}\frac{\partial F_{e}}{\partial v_{ek}}-V_{ej}\left( \mathbf{r},t\right)
\int\limits^{t}_{t_{0}}\frac{\partial V_{ek}\left( \mathbf{r},t_{1}\right) }{\partial r_{j}}dt_{1}\frac{\partial F_{e}}{\partial r_{ek}}
\nonumber
\\ 
&\displaystyle
+\frac{e}{m_{e}}\left(\nabla \varphi\left( \mathbf{r},t\right)\frac{\partial F_{e}}{\partial\mathbf{v}_{e}}
-\frac{1}{c}\Big[\mathbf{v}_{e}\times\mathbf{B}
_{0}\left(z,t \right) \Big]\right)\frac{\partial F_{e}}{\partial\mathbf{v}_{e}}
\nonumber
\\ 
&\displaystyle
-\left\{
\frac{\partial V_{ej}\left( \mathbf{r},t\right) }{\partial t}+
V_{ek}\left( \mathbf{r},t\right)\frac{\partial V_{ej}\left( \mathbf{r},t\right) }{\partial r_{k}}
\right.  
\nonumber
\\ 
&\displaystyle
\left. 
+ \frac{e}{m_{e}}\left(\mathbf{E}_{0y}\left(z,t \right) 
+\frac{1}{c}\Big[\mathbf{V}_{e}\left( \mathbf{r},t\right)\times\mathbf{B}_{0}\left(z,t \right) \Big]\right)_{j} 
\right\}
\nonumber
\\ 
&\displaystyle
\times\frac{\partial F_{e}\left(\mathbf{v}_{e},\mathbf{r}_{e},t \right)}{\partial v_{ej}}=0.
\label{8}
\end{eqnarray}
In the approximation of the spatially uniform RF field (i. e. for $\kappa =0$ in our case), the time dependent RF electric field is excluded from Eq. 
(\ref{8}) for the velocity  $\mathbf{V}_{e}$ for which the expression in brackets vanishes. In the case of the spatially inhomogeneous RF fields, this
selection of the velocity $\mathbf{V}_{e}$ provides the derivation of the solution for $F_{e}$ in the form of power series in the small parameter 
$\kappa
\delta r_{e}\ll 1$, where $\delta r_{e}$ is the amplitude of the displacement of electron in the RF field.
For RF fields  (\ref{1}) and (\ref{2}) this velocity is determined by the equations
\begin{eqnarray}
&\displaystyle 
\frac{\partial V_{ey}\left(z, t\right)}{\partial t}+
V_{ez}\left(z, t\right)\frac{\partial V_{ey}\left(z, t\right)}{\partial z}
\nonumber
\\ 
&\displaystyle
=-\frac{e}{m_{e}}\left(E_{0y}\left(z,t \right) 
+\frac{1}{c}V_{ez}\left(z, t\right)B_{0x}\left(z,t \right)\right),
\label{9}
\\
&\displaystyle 
\frac{\partial V_{ez}\left(z,t\right)}{\partial t}+
V_{ez}\left(z, t\right)\frac{\partial V_{ez}\left(z, t\right)}{\partial z}
\nonumber
\\ 
&\displaystyle
=\frac{e}{m_{e}c}V_{ey}\left(z, t\right)B_{0x}\left(z, t\right).
\label{10}
\end{eqnarray}
With new variables $z_{e}, t'$, determined by the relations\cite{Davidson}
\begin{eqnarray}
&\displaystyle 
z= z_{e}+\int\limits^{t'}_{0}V_{ez}\left(z_{e},t'_{1} \right) dt'_{1}, \quad t=t', 
\label{11}
\end{eqnarray}
Eqs. (\ref{9}) and (\ref{10}) becomes
\begin{eqnarray}
&\displaystyle 
\frac{\partial V_{ey}\left( z_{e},t'\right) }{\partial t'} =
-\frac{eE_{0y}}{m_{e}}e^{-\kappa \left(z_{e}+\int\limits^{t'}_{0}V_{ez}\left(z_{e},t'_{1} \right) dt'_{1}\right)}
\nonumber
\\ 
&\displaystyle
\times\left(\sin \omega_{0}t'+\frac{\kappa V_{ez}\left(z_{e},t' \right)}{\omega_{0}}\cos\omega_{0}t' \right), 
\label{12}
\end{eqnarray}
\begin{eqnarray}
&\displaystyle 
\frac{\partial V_{ez}\left( z_{e},t'\right) }{\partial t'} 
=\kappa\xi_{e0}e^{-\kappa \left(z_{e}+\int\limits^{t'}_{0}V_{ez}\left(z_{e},t'_{1} \right) dt'_{1}\right)}
\nonumber
\\ 
&\displaystyle
\times\omega_{0}V_{ey}\left(z_{e},t' \right)\cos\omega_{0}t',
\label{13}
\end{eqnarray}
where 
\begin{eqnarray}
&\displaystyle \xi_{e0}=\frac{eE_{0y}}{m_{e}\omega^{2}_{0}}
\label{14}
\end{eqnarray}
is the amplitude of the displacement of an electron along the coordinate $y$ at $z_{e}=0$.  For the collisionless plasma ${\text Re}\, \kappa^{-1}=L_{s}$ 
is the skin depth for the anomalous skin effect, 
\begin{eqnarray}
&\displaystyle L_{s}=\left(\frac{v_{Te}c^{2}}{\sqrt{\pi}\omega_{0}\omega_{pe}^{2}} \right)^{1/3}.
\label{15}
\end{eqnarray}
We will find the solutions for $V_{ey}\left(z_{e}, t\right)$ and $V_{ez}\left(z_{e}, t\right)$ in the form of the power series in the parameter
$\kappa\xi_{e0}\ll 1$. In this paper, we consider the case of the high frequency RF wave for which the RF electric field force acting on 
electrons in the skin layer  prevails over the RF Lorentz force. The procedure of the solution of system (\ref{12}), (\ref{13}) for the case of the low 
frequency RF wave, for which the RF Lorentz force dominates over the RF electric field force, is different and will be considered 
in the separate paper. It follows from Eq. (\ref{13}) that $V_{ez}$ is constant in zero-order approximation and without loss of the generality we put it to 
be equal to zero. In this approximation, we obtain from Eq. (\ref{12}) the equation for $V_{ey}$,
\begin{eqnarray}
&\displaystyle 
\frac{\partial V_{ey}\left(z_{e},t\right)}{\partial t}= -\frac{eE_{0y}\left(z_{e}\right)}{m_{e}}\sin \omega_{0}t,
\label{16}
\end{eqnarray}
with solution 
\begin{eqnarray}
&\displaystyle 
V_{ey}\left(z_{e},t\right)=\frac{eE_{0y}\left(z_{e}\right)}{m_{e}\omega_{0}}\cos \omega_{0}t
\nonumber
\\ 
&\displaystyle
=\frac{e}{cm_{\alpha}}A_{0y}\left(z_{e},t \right),
\label{17}
\end{eqnarray}
where $E_{0y}\left(z_{e}\right)=E_{0y}e^{-\kappa z_{e}}$ is the local value of the amplitude of the $\mathbf{E}_{0y}$ field. 

In the first order in $\kappa\xi_{e0}$, we find from Eq. (\ref{13}) the equation for $V_{ez}$,
\begin{eqnarray}
&\displaystyle 
\frac{\partial V_{ez}\left(z_{e},t\right)}{\partial t}= \frac{e\kappa E_{0y}\left(z_{e}\right)}{m_{e}\omega_{0}}
V_{ey}\left(z_{e},t\right)\cos \omega_{0}t
\nonumber
\\ 
&\displaystyle
= \kappa\frac{e}{m_{e}}\xi_{e}E_{0y}\left(z_{e}\right)
\cos^{2} \omega_{0}t,
\label{18}
\end{eqnarray}
where 
\begin{eqnarray}
&\displaystyle \xi_{e}=\xi_{e}\left(z_{e}\right)=\xi_{e0}e^{-\kappa z_{e}}
\label{19}
\end{eqnarray}
is the amplitude of the local displacement of electron along the coordinate $y$ at $z_{e}$. Equation (\ref{18}) is similar to the equation of 
the electron motion under the action of the ponderomotive force\cite{Schmidt} with solution
\begin{eqnarray}
&\displaystyle
V_{ez}\left(z_{e},t \right)=\kappa\xi_{e}\frac{e }{2m_{e}}E_{0y}\left(z_{e}\right)t
\nonumber
\\ 
&\displaystyle
+\frac{1}{4}\kappa\xi_{e}\frac{e }{m_{e}\omega_{0}}E_{0y}\left(z_{e}\right)\sin2\omega_{0}t.
\label{20}
\end{eqnarray}
With velocity $\mathbf{V}_{e}$ determined above, the Vlasov equation (\ref{8}) becomes
\begin{eqnarray}
&\displaystyle 
\frac{\partial F_{e}\left(\mathbf{v}_{e},\mathbf{r}_{e},t \right) }{\partial t}
+\mathbf{v}_{e}\frac{\partial F_{e}}{\partial\mathbf{r}_{e}}
+\frac{e}{m_{e}}\nabla \varphi \left(\mathbf{r}_{e}, t\right) \frac{\partial F_{e}}{\partial\mathbf{v}_{e}}
\nonumber
\\ 
&\displaystyle
+\kappa\xi_{e} v_{ez}\sin\omega_{0}t\frac{\partial F_{e}}{\partial y_{e}}
\nonumber
\\ 
&\displaystyle
+\kappa \xi_{e} v_{ey}\omega_{0}\cos\omega_{0}t\frac{\partial F_{e}}{\partial v_{ez}}=0.
\label{21}
\end{eqnarray}
Equation (\ref{21}) and the Vlasov equation for ions jointly with the Poisson equation (\ref{4}) for the potential 
$\varphi \left(\mathbf{r}_{e}, t\right)$ 
compose basic system of equations. It is important to note, that the spatial inhomogeneity and time dependence in the zero order in $\kappa\xi_{e}$ is 
excluded from the Maxwellian distribution (\ref{5}) in convective coordinates with velocity $V_{ey}$ determined by Eq. (\ref{17}). 
At the same time, the transition from $v_{z}$ to $v_{ez}$ introduces spatial inhomogeneity and time dependence of the first order in $\kappa\xi_{e}$ to 
$F_{e0}$. Therefore, the solution of the 
Vlasov equation (\ref{21}) for $F_{e0}\left(v_{ez},v_{ey},z_{e},t \right)$ may be presented in the form of power series in $\kappa\xi_{e}\ll 1$,
\begin{eqnarray}
&\displaystyle 
F_{e0}\left(v_{ez}, v_{ey}, z_{e}, t \right) =
F^{\left(0 \right)}_{e0}\left(v_{ez}, v_{ey}\right)
\nonumber
\\ 
&\displaystyle
+F^{\left(1 \right) }_{e0}\left(v_{ez}, v_{ey}, z_{e}, t \right),
\label{22}
\end{eqnarray}
where
\begin{eqnarray}
&\displaystyle 
F^{\left(0 \right) }_{e0}\left(v_{ez},v_{ey}\right) = \frac{n_{0e}}{2\pi v^{2}_{Te} }
\exp\left[-\frac{v^{2}_{ez}}{2v^{2}_{Te}}-\frac{v^{2}_{ey}}{2v^{2}_{Te}}\right]. 
\label{23}
\end{eqnarray}
With expansion (\ref{22}) the spatial inhomogeneity and time dependence of $F_{e0}\left(v_{ez}, v_{ey}, z_{e}, t \right)$ 
in the convective coordinates is determined by $F_{e0}^{\left(1 \right)}$, which is the solution of Eq. (\ref{21}) with 
$\varphi \left(\mathbf{v}_{e},\mathbf{r}_{e}, t\right) =0$,
\begin{eqnarray}
&\displaystyle 
\frac{\partial F_{e0}^{\left(1 \right)} }{\partial t}
+v_{ez}\frac{\partial F_{e0}^{\left(1 \right) } }{\partial z_{e}}
\nonumber
\\ 
&\displaystyle
=-\kappa\xi_{e}\omega_{0}\frac{v_{ey}}{2}\left(e^{-i\omega_{0}t} + e^{i\omega_{0}t}\right)
\frac{\partial F_{e0}^{\left(0 \right)}}{\partial v_{ez}}.
\label{24}
\end{eqnarray}
With new characteristic variable $z'_{e}=z_{e}-v_{ez}t$, the derivative over $z_{e}$ is excluded from Eq. (\ref{24}) and the solution to Eq. (\ref{24}) 
becomes
\begin{eqnarray}
&\displaystyle 
F_{e0}^{\left(1 \right)}\left(\mathbf{v}_{e},z',t \right) = -\kappa\xi_{e}\frac{v_{ey}}{2}\omega_{0}
\frac{\partial  F_{e0}^{\left(0 \right)} }{\partial v_{ez}}
\nonumber
\\ 
&\displaystyle
\times \left(\frac{e^{i\omega_{0} t}}{i\omega_{0} -\kappa v_{ez}}-\frac{e^{-i\omega_{0} t}}
{i\omega_{0} +\kappa v_{ez}} \right)+\Psi\left(v_{ez},v_{ey} \right).
\label{25}
\end{eqnarray}
The function $\Psi\left(v_{ez},v_{ey} \right)$ is determined by employing simple boundary conditions\cite{Shaing} determined for different values of  
coordinate $z_{e}$. The first condition is applied at $z_{e}=\infty$ for the electrons moving from  $z_{e}=\infty$ toward plasma boundary  $z_{e}=0$, i. e. 
for electrons with velocity $v_{ez}<0$. Because the electric field $E_{0y}\left(z_{e}\right)$ vanishes at $z_{e}=\infty$, the boundary condition $F_{e0}
^{\left(1\right)}\left(\mathbf{v}_{e},z'\rightarrow +\infty, t \right) =0$ determines $\Psi
\left(v_{ez}<0,v_{ey} \right)=0$ and  
\begin{eqnarray}
&\displaystyle 
F_{e0}^{\left(1 \right)}\left(v_{ey}, v_{ez}<0, z', t \right) =
-\kappa\xi_{e}\omega_{0}\frac{v_{ey}}{2}
\frac{\partial F_{e0}^{\left(0 \right)}}{\partial v_{ez}}
\nonumber
\\ 
&\displaystyle
\times \left(\frac{e^{i\omega_{0} t}}{i\omega_{0} -\kappa v_{ez}}
-\frac{e^{-i\omega_{0} t}}{i\omega_{0} +\kappa v_{ez}}
\right). 
\label{26}
\end{eqnarray}
The second boundary condition is the condition of the specular reflection of electrons  at the plasma boundary $z=0$,
\begin{eqnarray}
&\displaystyle 
F_{e0}^{\left(1 \right)}\left(v_{ey}, v_{ez}<0, z=0, t\right)
\nonumber
\\ 
&\displaystyle
= F_{e0}^{\left(1 \right)}\left(v_{ey}, v_{ez}>0, z=0, t\right).
\label{27}
\end{eqnarray}
This condition determines the solution for electron distribution function  $F_{e0}^{\left(1 \right)}\left(v_{ey}, v_{ez}>0,z',t \right)$ in a form
\begin{eqnarray}
&\displaystyle 
F_{e0}^{\left(1 \right)}\left(v_{ey}, v_{ez}>0,z',t \right) =
-\kappa\omega_{0}\frac{v_{ey}}{2}
\frac{\partial F_{e0}^{\left(0 \right)}}{\partial v_{ez}}
\nonumber
\\ 
&\displaystyle
\times
\left[\xi_{e}\left(\frac{e^{i\omega_{0} t}}{i\omega_{0} -\kappa v_{ez}}
-\frac{e^{-i\omega_{0} t }}{i\omega_{0} +\kappa v_{ez}}
\right)
e^{-\kappa v_{ez}t }\right.
\nonumber
\\ 
&\displaystyle
\left.
+\frac{4\kappa \xi_{e0} v_{ez}}{\omega_{0}^{2}+\kappa^{2} v_{ez}^{2}}\cos \omega_{0} t\right].  
\label{28}
\end{eqnarray}
With the equilibrium distribution function $F_{e0}$, determined by Eq. (\ref{22}), 
the Vlasov equation (\ref{21}) for the function $f_{e}$ becomes
\begin{eqnarray}
&\displaystyle 
\frac{\partial f_{e}\left(\mathbf{v}_{e},\mathbf{r}_{e},t \right) }{\partial t}
+\mathbf{v}_{e}\frac{\partial f_{e}}{\partial\mathbf{r}_{e}}+\kappa\xi_{e}v_{ez}\sin\omega_{0}t\frac{\partial f_{e}}{\partial y_{e}}
\nonumber
\\ 
&\displaystyle
+\kappa \xi_{e} v_{ey}\omega_{0}\cos\omega_{0}t\frac{\partial f_{e}}
{\partial v_{ez}}+\frac{e}{m_{e}}\nabla \varphi \left(\mathbf{r}_{e}, t
\right)
\nonumber
\\ 
&\displaystyle
\times
\frac{\partial }{\partial\mathbf{v}_{e}}\left( F^{\left(0 \right)}_{e0}
+F^{\left(1 \right) }_{e0}\left(v_{ez}, v_{ey}, z_{e}, t \right)\right) =0,
\label{29}
\end{eqnarray}
which contains the electrostatic potential $\varphi \left(\mathbf{r}_{e}, t\right)$ of the self-consistent 
respond of a plasma on the RF wave. The solution to Eq. (\ref{29}) may be found in the form of power series in $\kappa\xi_{e}\ll 1$.
In this paper, we obtain the solution to Eq. (\ref{29}) for $f_{e}$ and $f_{i}$ in the zero order in $\kappa\xi_{e}$ and use them 
in the Poisson equation for the potential $\varphi \left(\mathbf{r}_{e}, t\right)$. On this way, we obtain the basic equations 
of the theory of the parametric instabilities which may be developed in the inductively coupled plasma.

\section{Electron ocsillating mode}\label{sec3}

In the zero order in $\kappa\xi_{e}$, the equilibrium distribution functions $F_{e0,i0}$ in the convective 
coordinates are determined by the spatially inhomogeneous functions $F^{(0)}_{e0,i0}\left(\mathbf{v}_{e,i}\right)$,
 and the Vlasov equation 
(\ref{29}) for $f_{e}\left(\mathbf{v}_{e},\mathbf{r}_{e},t \right)$ 
and similar equation for $f_{i}\left(\mathbf{v}_{i},\mathbf{r}_{i},t \right)$ 
do not contain the RF electric field in their convective-oscillating frames. Therefore  equations for $f_{i}$ and $f_{e}$ 
will be the same as for the plasma without RF field,
\begin{eqnarray}
&\displaystyle 
\frac{\partial f_{i}}{\partial
t}+\mathbf{v}_{i}\frac{\partial f_{i}}
{\partial\mathbf{r}_{i}}-\frac{e_{i}}{m_{i}}\nabla 
\varphi_{i}\left(\mathbf{r}_{i},t\right)\frac{\partial
F_{i0}}{\partial\mathbf{v}_{i}}=0,
\label{30}
\\
&\displaystyle \frac{\partial f_{e}}{\partial
t}+\mathbf{v}_{e}\frac{\partial f_{e}}
{\partial\mathbf{r}_{e}}-\frac{e}{m_{e}}\nabla 
\varphi_{e}\left(\mathbf{r}_{e},t\right)\frac{\partial
F_{e0}}{\partial\mathbf{v}_{e}}=0.
\label{31}
\end{eqnarray}
The solution of the linearised equations for $f_{i}$ Fourier transformed over $\mathbf{r}_{i}$ is
\begin{eqnarray}
&\displaystyle 
f_{i}\left(\mathbf{v}_{i}, \mathbf{k}_{i}, t\right)=i\frac{e_{i}}{m_{i}}\mathbf{k}_{i}\frac{\partial F_{i0}}{\partial \mathbf{v}_{i}}\int 
\limits^{t}_{0}dt_{1}\varphi_{i}\left(\mathbf{k}_{i}, t_{1}\right)
\nonumber
\\ 
&\displaystyle
\times
e^{-i\mathbf{k}_{i}\mathbf{v}_{i}\left(t-t_{1}\right)},
\label{32}
\end{eqnarray}
where $\varphi_{i}\left(\mathbf{k}_{i}, t_{1}\right)$ is the Fourier transform of the potential 
$\varphi_{i}\left(\mathbf{r}_{i}, t_{1}\right)$ over $\mathbf{r}_{i}$,
\begin{eqnarray}
&\displaystyle 
\varphi_{i}\left(\mathbf{k}_{i}, t_{1}\right)=\frac{1}{\left(2\pi\right)^{3}}\int 
d\mathbf{r}_{i}\varphi_{i}\left(\mathbf{r}_{i}, t_{1}\right)e^{-i\mathbf{k}_{i}\mathbf{r}_{i}}.
\label{33}
\end{eqnarray}
The ion density perturbation $n_{i}\left(\mathbf{k}_{i}, t\right)$ Fourier transformed over $\mathbf{r}_{i}$ with the conjugate wave vector
$\mathbf{k}_{i}$  is
\begin{eqnarray}
&\displaystyle 
n_{i}\left(\mathbf{k}_{i}, t\right)=\int f_{i}\left(\mathbf{v}_{i}, \mathbf{k}_{i}, t\right)d\mathbf{v}_{i}
=i\frac{e_{i}}{m_{i}}\mathbf{k}_{i}
\nonumber
\\ 
&\displaystyle
\times\int d\mathbf{v}_{i} \frac{\partial F_{i0}}{\partial \mathbf{v}_{i}}\int 
\limits^{t}_{0}dt_{1}\varphi_{i}\left(\mathbf{k}_{i}, t_{1}\right)e^{-i\mathbf{k}_{i}\mathbf{v}_{i}\left(t-t_{1}\right)}.
\label{34}
\end{eqnarray}
The Fourier transform $n_{e}\left(\mathbf{k}_{e}, t\right)$ of the electron density 
perturbation performed in the electron frame is given 
by equation
\begin{eqnarray}
&\displaystyle 
n_{e}\left(\mathbf{k}_{e}, t\right)=\int f_{e}\left(\mathbf{v}_{e}, \mathbf{k}_{e}, t\right)d\mathbf{v}_{e}
=i\frac{e}{m_{e}}\mathbf{k}_{e}
\nonumber
\\ 
&\displaystyle
\times\int d\mathbf{v}_{e} \frac{\partial F_{e0}}{\partial \mathbf{v}_{e}}\int 
\limits^{t}_{0}dt_{1}\varphi_{e}\left(\mathbf{k}_{e}, t_{1}\right)e^{-i\mathbf{k}_{e}\mathbf{v}_{e}\left(t-t_{1}\right)}.
\label{35}
\end{eqnarray}
which is the same as Eq. (\ref{34}) for $n_{i}\left(\mathbf{k}_{i}, t\right)$ with changing ion on electron subscripts.

The perturbations  of the ion, (\ref{34}), and electron, (\ref{35}), density are used in the Poisson equation (\ref{4})
which may be the equation for $\varphi_{i}\left(\mathbf{k}_{i}, t_{1}\right)$ by the Fourier transform of Eq. (\ref{4}) over $\mathbf{r}_{i}$,
\begin{eqnarray}
& \displaystyle 
k^{2}_{i}\varphi_{i}\left(\mathbf{k}_{i},t\right)=4\pi e\left(n_{i}\left(\mathbf{k}_{i}, t\right)\right.
\nonumber
\\ 
&\displaystyle
\left.-\int d\mathbf{r}_{i}n_{e}\left(\mathbf{r}_{e}, t\right)e^{-i\mathbf{k}_{i}\mathbf{r}_{i}}\right),
\label{36}
\end{eqnarray}
or as the equation for  $\varphi_{e}\left(\mathbf{k}_{e},t\right)$ by the Fourier transform of Eq. (\ref{4}) over $\mathbf{r}_{e}$. 
For the deriving the Poisson equation for $\varphi_{i}\left(\mathbf{k}_{i} , t\right)$ the Fourier transforms 
$n^{(i)}_{e}\left(\mathbf{k}_{i}, t\right)$ and 
$\varphi^{(i)}_{e}\left(\mathbf{k}_{i}, t\right)$ of $n_{e}\left(\mathbf{r}_{e}, t\right)$ 
and $\varphi_{e}\left(\mathbf{r}_{e}, t\right)$ 
over $\mathbf{r}_{i}$ should be determined. Using Eq. (\ref{7}), which determines the relations among the coordinates in the 
laboratory, ion and electron frames, we find that the electron density perturbation $n_{e}\left(\mathbf{r}_{e}, t\right)$ 
Fourier transformed over $\mathbf{r}_{i}$ is
\begin{eqnarray}
&\displaystyle
n^{(i)}_{e}\left(\mathbf{k}_{i}, t\right)=\int d\mathbf{r}_{i}n_{e}\left(\mathbf{r}_{e}, t\right)e^{-i\mathbf{k}_{i}\mathbf{r}_{i}}=\int d
\mathbf{r}_{e}n_{e}\left(\mathbf{r}_{e}, t\right)
\nonumber
\\ 
&\displaystyle
\times 
\exp\left(-i\mathbf{k}_{i}\mathbf{r}_{e}-ik_{iy}\int dt_{1}\left(V_{ey}\left(t_{1}\right)-V_{iy}\left(t_{1}\right)\right)\right.
\nonumber
\\ 
&\displaystyle
\left.-ik_{iz}\int dt_{1}\left(V_{ez}\left(t_{1}\right)-V_{iz}\left(t_{1}\right)\right)\right),
\label{37}
\end{eqnarray}
where velocities $V_{ey}\left(t_{1}\right)$ and $V_{ez}\left(t_{1}\right)$ are determined by Eqs. (\ref{17}) and (\ref{20}). The velocities $V_{iy}
\left(t_{1}\right)$ and $V_{iz}\left(t_{1}\right)$, which are determined by the same Eqs. (\ref{17}) and (\ref{20}) with subscript $i$ 
instead of $e$, are in $m_{i}/m_{e}$  times less than $V_{ey}$ and $V_{ez}$ and are neglected in what follows. 
One comment should be made concerning the uniformly accelerated 
part of the velocity $V_{ez}\left(t\right)$, resulted from the action of the ponderomotive force on electrons. 
It is clear that velocity  $V_{ez}\left(t\right)$ can't grow infinitely. 
After the development of the parametric instabilities, we must take into account 
the deceleration of the electrons due to 
their scattering by the turbulent electric fields powered by the parametric instabilities. 
In the steady state, determined by the relation 
\begin{eqnarray}
&\displaystyle
\kappa\xi_{e}\frac{e E_{0y}}{2m_{e}}-\nu_{eff}U_{0z}=0,
\label{38}
\end{eqnarray}
where $\nu_{eff}$ is the 'effective collision frequency' of the electrons with plasma turbulence, the  $V_{ez}\left(t\right)$ 
velocity at time $t\gg \nu^{-1}_{eff}$ is determined by the relation
\begin{eqnarray}
&\displaystyle
V_{ez}\left(t\right)= U_{0z}+\frac{1}{4}\kappa\xi_{e}\frac{e E_{0y}}{m_{e}\omega_{0}}\sin2\omega_{0}t.
\label{39}
\end{eqnarray}
With velocities $V_{ey}\left(t\right)$ and $V_{ez}\left(t\right)$ determined by Eqs. (\ref{17}) and (\ref{39}) relation (\ref{37}) becomes
\begin{eqnarray}
&\displaystyle 
n^{(i)}_{e}\left(\mathbf{k}_{i}, t\right)=n^{(e)}_{e}\left(\mathbf{k}_{i}, t\right)
\exp\Big(-ik_{iy}\xi_{e}\sin\omega_{0}t
\nonumber
\\ 
&\displaystyle
-ik_{iz}U_{ez}t +ik_{iz}\eta_{e}\cos 2\omega_{0}t\Big)
\nonumber
\\ 
&\displaystyle
=\sum\limits_{n=-\infty}^{\infty}\sum\limits^{\infty}_{m=-\infty}J_{n}\left(k_{iy}\xi_{e}\right)J_{m}\left(k_{iz}\eta_{e}\right)
\nonumber
\\ 
&\displaystyle
\times
e^{i\left(n+2m\right)\omega_{0}t-ik_{iz}U_{ez}t+i\frac{m\pi}{2}}n^{(e)}_{e}\left(\mathbf{k}_{i}, t\right),
\label{40}
\end{eqnarray}

where $\xi_{e}$ is determined by Eq. (\ref{19}) and 
\begin{eqnarray}
&\displaystyle 
\eta_{e}=\eta_{e}\left(z_{e}\right)=\frac{1}{8}\kappa\xi^{2}_{e}\left(z_{e}\right)
\label{41}
\end{eqnarray}
is the amplitude of the electron  displacement in the RF electric field along coordinate $z_{e}$. In Eq. (\ref{40}), 
$J_{m}\left(x\right)$  is first kind Bessel function of order $m$  and the relations
\begin{eqnarray}
&\displaystyle
e^{ik_{iy}\xi_{e}\sin \omega_{0}t}=\sum\limits ^{\infty}_{n=-\infty}J_{n}\left(k_{iy}\xi_{e}
\right)e^{in\omega_{0}t},\label{42}
\end{eqnarray}
\begin{eqnarray}
&\displaystyle
e^{ik_{iz}\eta_{e}\cos 2\omega_{0}t}=\sum\limits^{\infty}_{m=-\infty}J_{m}\left(k_{iz}
\eta_{e}\right)e^{im\left(2\omega_{0}t+\frac{\pi}{2}\right)}
\label{43}
\end{eqnarray}
were used. It follows from Eq. (\ref{40}) that the perturbation of the electron density 
$n^{(e)}_{e}\left(\mathbf{k}_{i}\right)e^{-i\omega t}$, determined in the oscillating 
electron frame -- the electron oscillating mode, is detected in the laboratory (ion) frame as 
the sum of this mode with infinite number of the harmonics, 
$\sim e^{-i\left(\omega-\left(n+2m\right)\omega_{0}-k_{iz}U_{ez}\right)t}$ with Doppler 
shifted frequency.  

The relation between the Fourier transform $\varphi_{e}\left( \mathbf{k}_{e}, t\right)$ of 
the potential $\varphi_{e}\left( \mathbf{r}_{e}, t\right)$ over 
$\mathbf{r}_{e}$, involved in the expression 
for $n_{e}\left(\mathbf{r}_{e}, t\right)$, and the Fourier transform $\varphi_{i}\left( 
\mathbf{k}_{i}, t\right)$ of the potential $\varphi_{e}\left( 
\mathbf{r}_{i}, t\right)$ over $\mathbf{r}_{i}$, when it is 
used in $n^{(i)}_{e}\left(\mathbf{k}_{i}, t\right)$, is derived similar and is determined by 
the relation
\begin{eqnarray}
&\displaystyle 
\varphi^{(e)}_{e}\left(\mathbf{k}_{e},t_{1}\right)=\exp\Big(ik_{iy}\xi_{e}
\sin \omega_{0}t_{1}+ik_{iz}U_{ez}t_{1}
\nonumber
\\ 
&\displaystyle
-ik_{iz}\eta_{e}\cos 
2\omega_{0}t_{1}\Big)\varphi_{i}\left(\mathbf{k}_{i}, t_{1}\right),
\label{44}
\end{eqnarray}
which follows from the identity $\varphi_{e}\left(\mathbf{r}_{e},t_{1}\right)=\varphi_{i}
\left(\mathbf{r}_{i},t_{1}\right)$, and relation (\ref{40}). In Eq. (\ref{44}), we employ the 
local approximation for the electric field $E_{0y}$: because of the small amplitude of the 
electron oscillation in RF field along coordinate $z_{e}$, $E_{0y}$   means the local value 
$E_{0y}e^{-\kappa z_{e}}$  of the weakly inhomogeneous electric field $E_{y}$. 

With Eqs. (\ref{40}) and (\ref{44}) employed in Eq. (\ref{39})  for $n_{e}^{(i)}
\left(\mathbf{r}_{i},t\right)$   the Poisson equation (\ref{36}), Fourier 
transformed over time,  gives the equation
\begin{eqnarray}
&\displaystyle
\left(1+\varepsilon_{i}\left(\mathbf{k}_{i}, \omega\right)\right)\varphi_{i}\left(\mathbf{k}
_{i}, \omega\right)
\nonumber
\\ 
&\displaystyle
+\sum\limits_{n=-\infty}^{\infty}\sum
\limits^{\infty}_{p=-\infty}\sum\limits_{r=-\infty}^{\infty}\sum\limits_{q=-\infty}^{\infty}
J_{n-2r}\left(k_{iy}\xi_{e}\right)
\nonumber
\\ 
&\displaystyle
\times 
J_{n-2r-q+p}\left(k_{iy}
\xi_{e}\right)
J_{r}\left(k_{iz}\eta_{e}\right)
J_{r-q}\left(k_{iz}\eta_{e}\right)
\nonumber
\\ 
&\displaystyle
\times 
e^{i\frac{\pi q}{2}}
\varepsilon_{e}\left(\mathbf{k}_{i}, \omega-k_{iz}U_{ez}-n\omega_{0}\right)
\nonumber
\\ 
&\displaystyle
\times 
\varphi_{i}
\left(\mathbf{k}_{i}, \omega+p\omega_{0}\right)=0,
\label{45}
\end{eqnarray}
which determines  the evolution of the  electrostatic potential $\varphi_{i}\left(\mathbf{k}
_{i}, \omega\right)$ in the skin layer of an inductively coupled plasma.
In  Equation (\ref{45}), $\varepsilon_{i, e}\left(\mathbf{k}_{i}, \omega\right)$ is the ion 
(electron) dielectric permittivity. For the Maxwellian distribution $F_{0\alpha}
\left(\mathbf{v}_{\alpha}\right)$,
\begin{eqnarray}
F_{0\alpha}\left(\mathbf{v}_{\alpha}\right)=\frac{n_{0}}{\left(2\pi v^{2}_{Ti}\right)^{3/2}}
\exp\left(-\frac{v_{\alpha}^{2}}{2v^{2}_{Ti}}\right),
\label{46}
\end{eqnarray}
dielectric permittivity $\varepsilon_{\alpha}$ $\left(\alpha= i, e\right)$ is equal to
\begin{eqnarray}
\varepsilon_{\alpha}\left(\mathbf{k}_{i}, \omega\right)=\frac{1}{k^{2}_{i}\lambda^{2}_{D
\alpha}}\left(1+i\sqrt{\pi}z_{\alpha}
W\left(z_{\alpha}\right)\right),
\label{47}
\end{eqnarray}
where $\lambda_{D\alpha}$ is the Debye radius, $z_{\alpha} =\omega/\sqrt{2}k_{i}v_{T\alpha}$, 
$W\left(z\right)=e^{- z^{2}}\left(1 +\left(2i / \sqrt {\pi}
\right)\int\limits_{0}^{z}
e^{t^{2}}dt \right)$ is the the complex error function.

\bigskip
\section{The parametric instability of the skin layer driven by the ponderomotive current }
\label{sec4}

For numerical solution of Eq. (\ref{45}) we present this equation in a form of the infinite 
system of equations for the fundamental mode $\varphi_{i}
\left(\mathbf{k}_{i}, \omega\right)$ and harmonics $\varphi_{i}\left(\mathbf{k}_{i}, \omega-m
\omega_{0}\right)$.  By replacing  $\omega$ 
on $\omega-m\omega_{0}$ in Eq. (\ref{45}), where $m$ is an integer,  we find 
\begin{eqnarray}
&\displaystyle
\left(1+\varepsilon_{i}\left(\mathbf{k}_{i}, \omega-m\omega_{0}\right)\right)
\varphi_{i}\left(\mathbf{k}_{i}, \omega-m\omega_{0}\right)
\nonumber
\\ 
&\displaystyle
+\sum\limits_{r=-\infty}^{\infty}\sum\limits^{\infty}_{n=-\infty}\sum\limits_{t=-\infty}^{\infty}\sum\limits_{q=-\infty}^{\infty}
J_{r+m+2t}\left(k_{iy}\xi_{e}\right)
\nonumber
\\ 
&\displaystyle
\times 
J_{r+2t+q+n}\left(k_{iy}\xi_{e}\right)
J_{t}\left(k_{iz}\eta_{e}\right)J_{t-q}\left(k_{iz}\eta_{e}\right)
\nonumber
\\ 
&\displaystyle
\times 
\left(-1\right)^{m+n-q}e^{i\frac{\pi q}{2}}
 \varepsilon_{e}\left(\mathbf{k}_{i}, \omega-k_{iz}U_{ez}+r\omega_{0}\right)
\nonumber
\\ 
&\displaystyle
\times  
 \varphi_{i}\left(\mathbf{k}_{i}, \omega-n\omega_{0}\right)=0.
\label{48}
\end{eqnarray}

Eq. (\ref{48}) forms the infinite system of equations 
\begin{eqnarray}
&\displaystyle
\sum\limits_{n=-\infty}^{\infty}a_{mn}\varphi_{i}\left(\mathbf{k}_{i}, \omega-n\omega_{0}\right)=0,
\label{49}
\end{eqnarray}
where $n$ and $m$ are integer numbers and the coefficients $a_{mn}$ are determined by relation
\begin{eqnarray}
&\displaystyle
a_{mn}=\delta_{mn}+\left(1+\varepsilon_{i}\left(\mathbf{k}_{i}, \omega-m\omega_{0}\right)\right)^{-1}
\nonumber
\\ 
&\displaystyle
\times 
\sum\limits_{r=-\infty}^{\infty}\sum\limits_{t=-\infty}^{\infty}\sum\limits_{q=-\infty}^{\infty}e^{i\left(m+n-\frac{q}{2}\right)\pi}
\nonumber
\\ 
&\displaystyle
\times 
J_{r+m+2t}\left(k_{iy}\xi_{e}\right)J_{r+2t+q+n}\left(k_{iy}\xi_{e}\right)J_{t}\left(k_{iz}\eta_{e}\right)
\nonumber
\\ 
&\displaystyle
\times 
J_{t-q}\left(k_{iz}\eta_{e}\right)
\varepsilon_{e}\left(\mathbf{k}_{i}, \omega-k_{iz}U_{ez}+r\omega_{0}\right).
\label{50}
\end{eqnarray}
The equality to zero of the determinant of this homogeneous system, 
\begin{eqnarray}
&\displaystyle
\text{det}\parallel a_{mn}\parallel=0,
\label{51}
\end{eqnarray}
gives the dispersion equation for system (\ref{49}). Below we present the numerical solution of this dispersion equation for 
system (\ref{49}) limited by three equations: for the potential $\varphi_{i}\left(\mathbf{k}_{i}, \omega\right)$ 
and its harmonics $\varphi_{i}\left(\mathbf{k}_{i}, \omega-\omega_{0}\right)$ and $\varphi_{i}\left(\mathbf{k}_{i}, \omega+\omega_{0}\right)$, 
i. e. for $m=0, \pm 1$ and $n= 0, \pm 1$. The summation indexes $r$, $t$, $q$ in coefficients $a_{mn}$ were limited by the interval $[-10, 10]$. In the 
numerical solution of Eq. (\ref{51})  we use the normalized frequencies $\hat{\omega}=\omega/\omega_{pe}$, $\hat{\omega}_{0}=\omega_{0}/\omega_{pe}$,  
and $\hat{\nu}_{eff}=\nu_{eff}/\omega_{pe}$, the normalized 
electric field $\hat{E}_{0y}=E_{0y}/\sqrt{4\pi n_{0e}T_{e}}$, and the normalized velocity of the ponderomotive current 
$\hat{U}_{ez}=U_{ez}/v_{Te}$. The Bessel functions arguments $k_{iy}\xi_{e}$ and $k_{iz}
\eta_{e}$ in the normalized variables are
\begin{eqnarray}
&\displaystyle
k_{iy}\xi_{e}=k_{iy}
\lambda_{De}\frac{\hat{E}_{0y}}{\hat{\omega}^{2}_{0}},  
\nonumber
\\ 
&\displaystyle
k_{iz}\eta_{e}=k_{iz}\lambda_{De}
\frac{\kappa\lambda_{De}}{8}
\left( \frac{\hat{E}_{0y}}{\hat{\omega}^{2}_{0}}\right) ^{2},
\label{52}
\end{eqnarray}
and
\begin{eqnarray}
&\displaystyle
k_{iz}\hat{U}_{ez}=k_{iz}\lambda_{De}
\frac{\kappa\lambda_{De}}{2\hat{\nu}_{eff}}\left( \frac{\hat{E}_{0y}}{\hat{\omega}_{0}}
\right) ^{2}.
\label{53}
\end{eqnarray}

\begin{figure}[!htbp]
\includegraphics[width=0.4\textwidth]{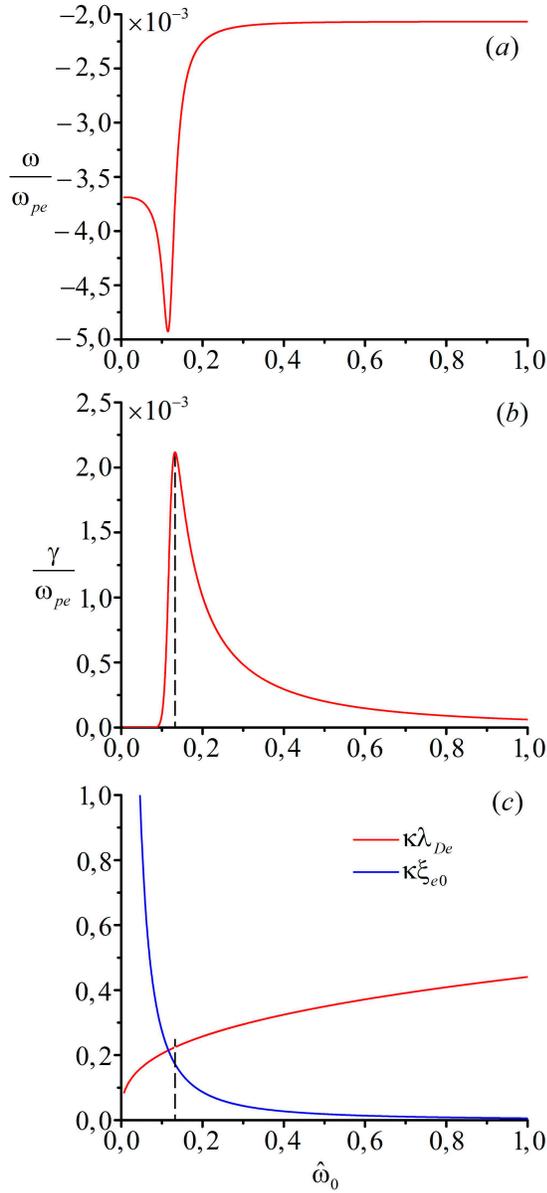}
\caption{\label{fig1} The normalized frequency $\omega/\omega_{pe}$ (a), normalized growth rate 
$\gamma/\omega_{pe}$ (b) and  $\kappa\lambda_{De}$ (red line) and $\kappa\xi_{e0}$ (blue line) plots
versus normalized frequency $\hat{\omega}_{0}$ for $\hat{E}_{0y} = 0.0134$.}
\end{figure}

\begin{figure}[!htbp]
\includegraphics[width=0.4\textwidth]{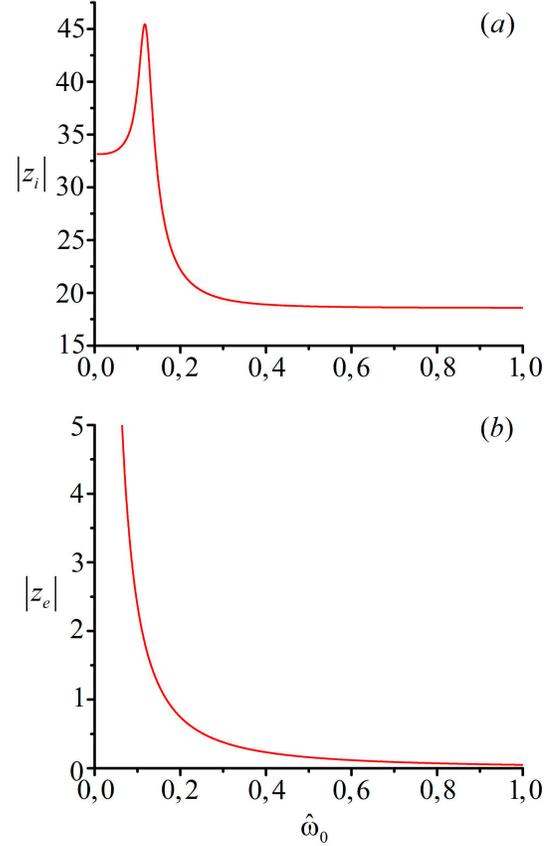}
\caption{\label{fig1a} The $\left|z_{i} \right|$ and $\left|z_{e} \right|$ parameters
versus normalized frequency $\hat{\omega}_{0}$ for $\hat{E}_{0y} = 0.0134$.}
\end{figure}

\begin{figure}[!htbp]
\includegraphics[width=0.4\textwidth]{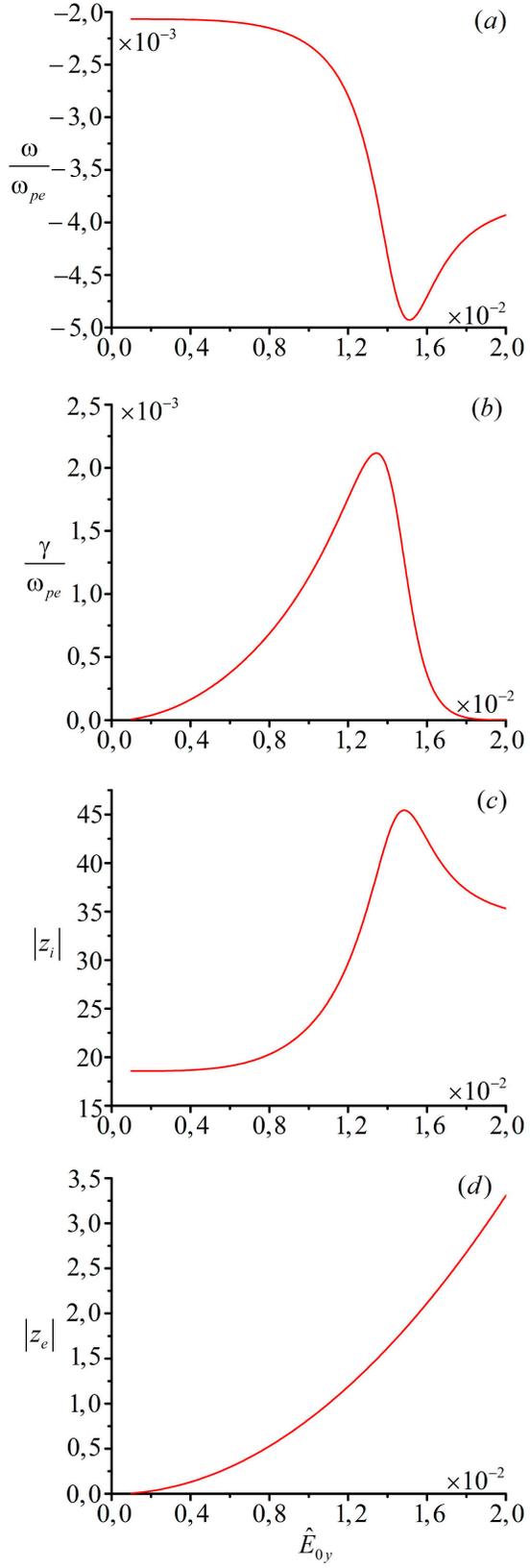}
\caption{\label{fig2} The normalized frequency $\omega/\omega_{pe}$, normalized growth rate $
\gamma/\omega_{pe}$, $\left|z_{i} \right|$ and $\left|z_{e} \right|$
versus $\hat{E}_{0y}$ for $\hat{\omega}_{0} = 0.132$.}
\end{figure}

\begin{figure}[!htbp]
\includegraphics[width=0.4\textwidth]{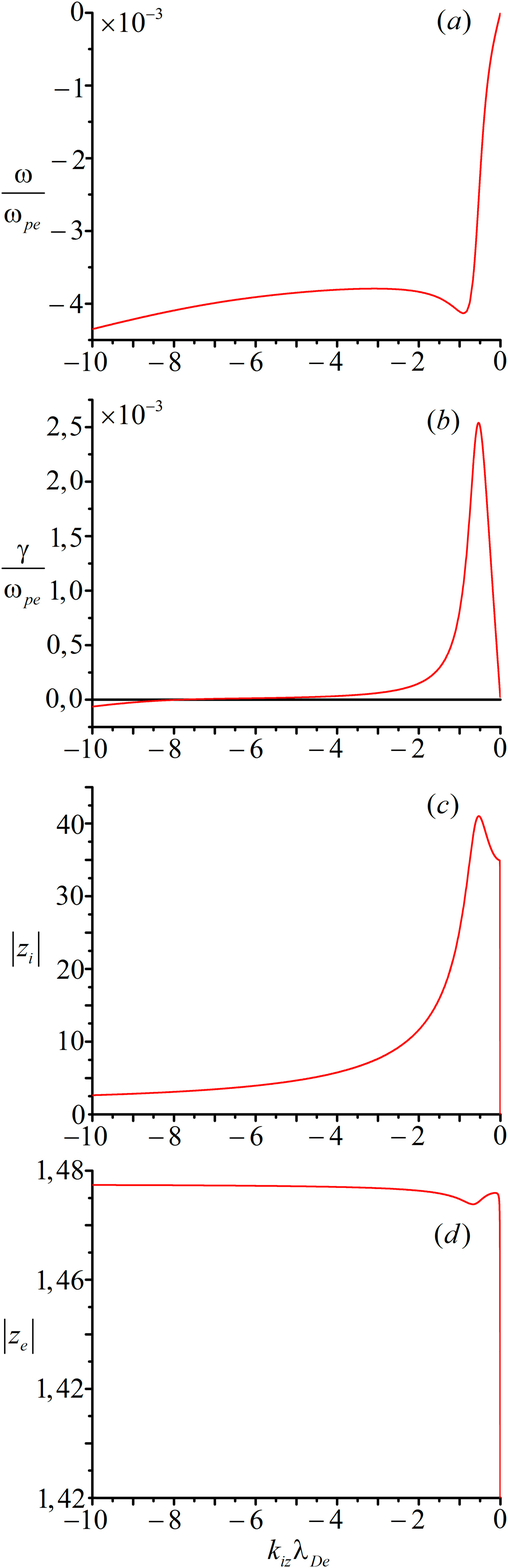}
\caption{\label{fig3} The normalized frequency $\omega/\omega_{pe}$, normalized growth rate 
$\gamma/\omega_{pe}$, $\left|z_{i} \right|$ and $\left|z_{e} \right|$
versus $k_{iz}\lambda_{De}$.}
\end{figure}

\begin{figure}[!htbp]
\includegraphics[width=0.4\textwidth]{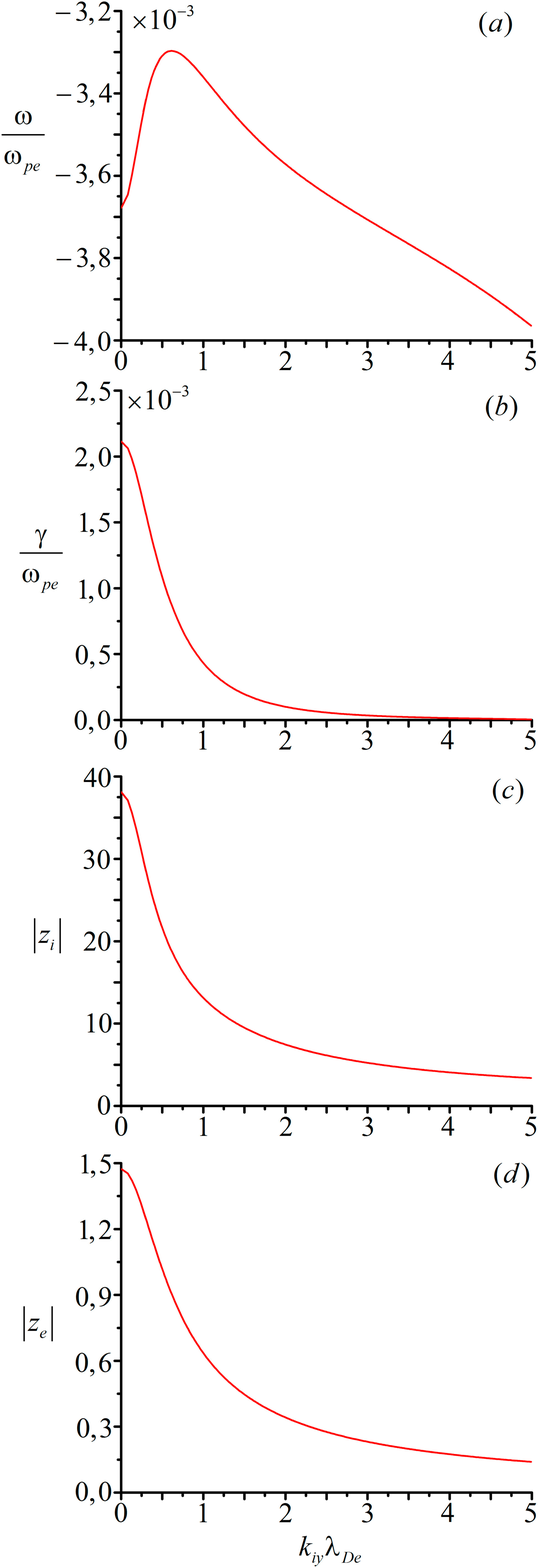}
\caption{\label{fig4} The normalized frequency $\omega/\omega_{pe}$, normalized growth rate $
\gamma/\omega_{pe}$, $\left|z_{i} \right|$ and $\left|z_{e} \right|$
versus $k_{iy}\lambda_{De}$.}
\end{figure}

\begin{figure}[!htbp]
\includegraphics[width=0.4\textwidth]{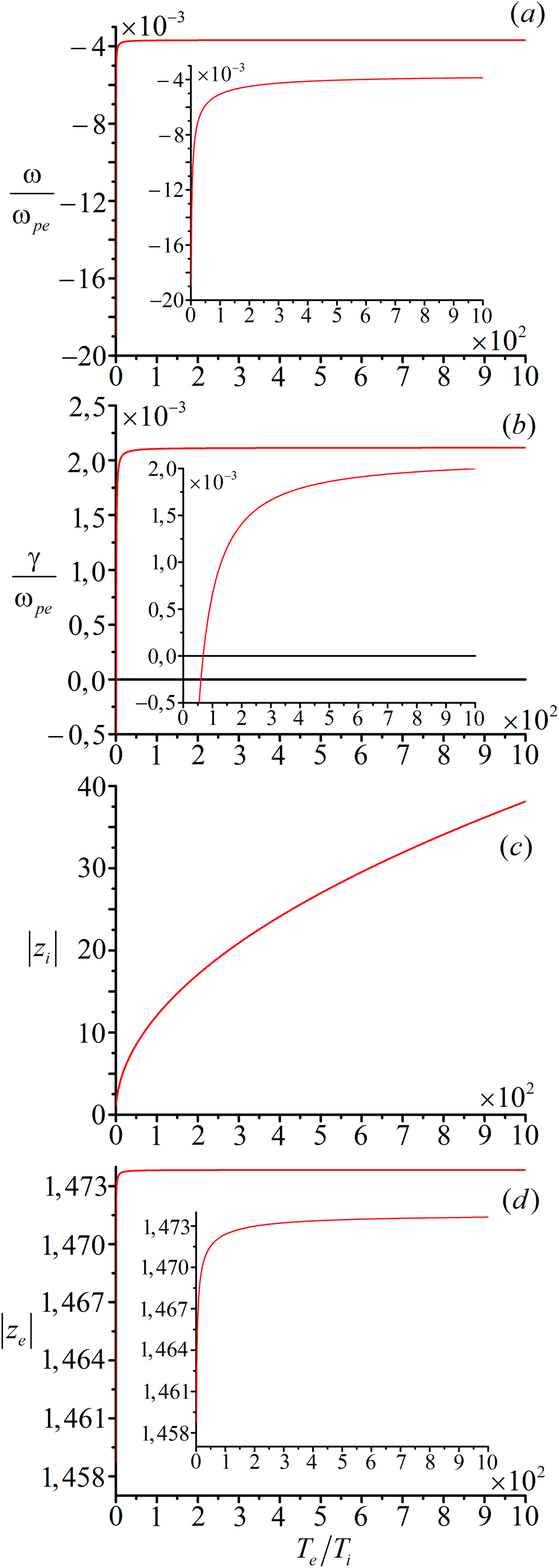}
\caption{\label{fig5} The normalized frequency $\omega/\omega_{pe}$, normalized growth rate $
\gamma/\omega_{pe}$, $\left|z_{i} \right|$ and $\left|z_{e} \right|$
versus $T_{e}/T_{i}$.}
\end{figure}

\begin{figure}[!htbp]
\includegraphics[width=0.4\textwidth]{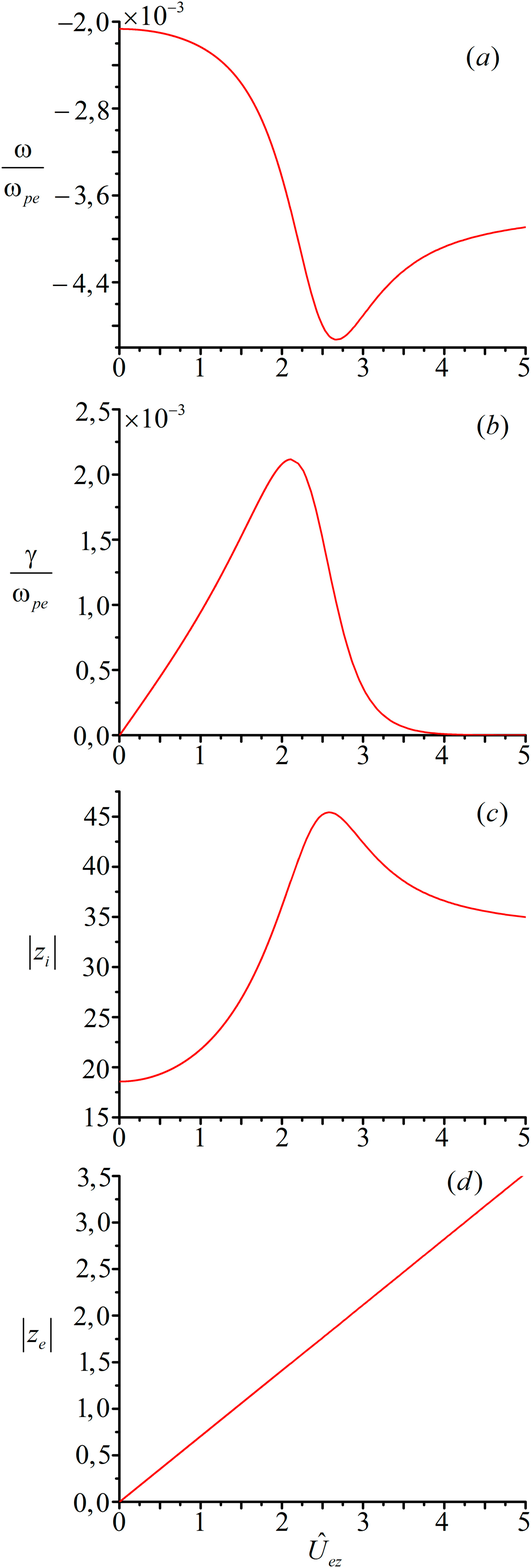}
\caption{\label{fig6} The normalized frequency $\omega/\omega_{pe}$, normalized growth rate $
\gamma/\omega_{pe}$, $\left|z_{i} \right|$ and $\left|z_{e} \right|$
versus $\hat{U}_{ez}$.}
\end{figure}

\begin{figure}[!htbp]
\includegraphics[width=0.4\textwidth]{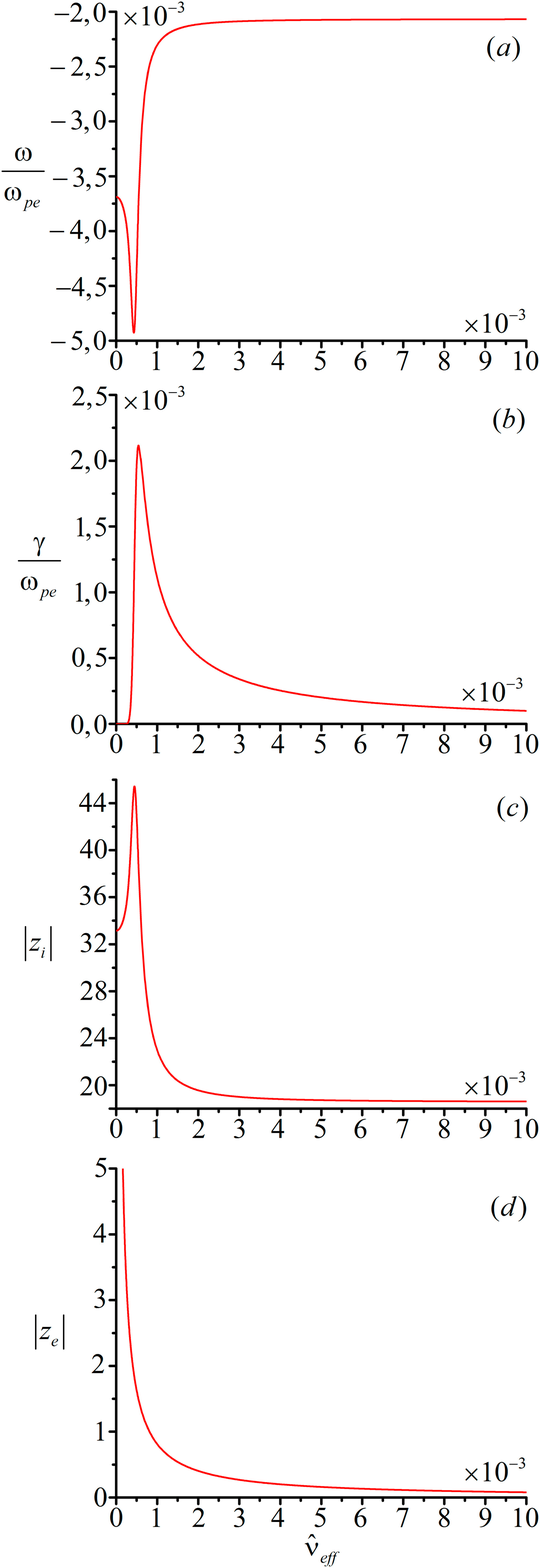}
\caption{\label{fig7} The normalized frequency $\omega/\omega_{pe}$, normalized growth rate $
\gamma/\omega_{pe}$, $\left|z_{i} \right|$ and $\left|z_{e} \right|$
versus $\hat{\nu}_{eff}$.}
\end{figure}

The results of the numerical solutions are presented in Figs. \ref{fig1}--\ref{fig7}. 
In Fig. \ref{fig1},  the solutions for $\omega/\omega_{pe}$, $\gamma/\omega_{pe}$, and in Fig. \ref{fig1a}, the solutions for  $z_{i}$ 
and $z_{e}$ are presented versus the normalized frequency 
$\hat{\omega}_{0}$ for the normalized electric field $\hat{E}_{0y} = 0.0134$. For a plasma with electron temperature $T_{e}=4 eV$ and density $n_{0e}
=10^{12}cm^{-3}$ this dimensionless values of $E_{0y}$ correspond to $E_{0y}=37\, V/cm$. 
The magnitudes of other parameters employed in both these figures, as well as in Fig. \ref{fig2}, are: $T_{e}/T_{i} = 10^{3}$, 
$k_{iy}\lambda_{De} = 10^{-3}$, $\kappa\lambda_{De} = 0.26$, $k_{iz}\lambda_{De} = -0.675$, 
$\hat{\nu}_{eff} = 5.5\cdot 10^{-4}$, $m_{i}/m_{e} = 40\cdot1840$ (Ar). Figs. \ref{fig1} and \ref{fig1a}
demonstrate the existence the instability of the kinetic type which develops due to the inverse electron Landau damping ($|z_{i}|\gg 1$, $|z_{e}|\sim 1$) 
with negative normalized frequency and with the growth rate comparable with frequency. 
It follows from Fig. \ref{fig1} that the growth rate maximum attains at $\hat{\omega}_{0} = 0.132$ for $\hat{E}_{0y}=0.0134$. 
Figure \ref{fig1}(c) displays that for these values of $\hat{\omega}_{0}$ and $\hat{E}_{0y}$, $\xi_{e0}\sim \lambda_{De}$ for the  growth rate maximum. 
This value of $\hat{\omega}_{0}$ is used in Fig. \ref{fig2} where solutions for $\omega/\omega_{pe}$, $\gamma/
\omega_{pe} $, $z_{i}$ and $z_{e}$ versus $\hat{E}_{0y}$ are presented. Note, that in Fig. 
\ref{fig2}, as well as in Figs. \ref{fig3}--\ref{fig7}, the solution for the 
normalized frequency $\omega/\omega_{pe}$ is presented in panel (a), the normalized growth rate $\gamma/\omega_{pe}$ is presented in panel (b), 
and the arguments $z_{i}$ and $z_{e}$ of the $W$-functions in $\varepsilon_{i}$ and $\varepsilon_{e}$ 
are presented in panels (c) and (d) respectively. 

The dependences of the $\omega/\omega_{pe}$, $\gamma/\omega_{pe}$, $z_{i}$ and $z_{e}$ on the 
normalized wavenumbers $k_{z}\lambda_{De}$ and $k_{y}\lambda_{De}$ are presented 
in Fig. \ref{fig3} and Fig. \ref{fig4} respectively. 
Figure \ref{fig4} displays that the wave vector $k_{z}$ is directed to the plasma boundary 
($k_{z}$ is negative) and the growth rate has maximum value at 
$k_{z}\lambda_{De}=-0.675$. Therefore we use $k_{z}\lambda_{De}=-0.675$ and $k_{y}\lambda_{De}=10^{-3}$ 
in our numerical calculations. 

The solutions for $\omega/\omega_{pe}$, $\gamma/\omega_{pe}$, $z_{i}$ and $z_{e}$ versus the 
$T_{e}/T_{i}$ ratio are presented in Fig. \ref{fig5}.
Figure \ref{fig5} displays that the growth rate maximum attains at $T_{e}/T_{i}\sim 10^{2}$ 
and this maximum holds up to $T_{e}/T_{i}\sim 10^{3}$ and 
above. The detected instability exists in plasmas even with hot ions for which $T_{i}=T_{e}$, 
however with the growth rate in 10 times less than maximum value. The instability is absent 
in plasmas where the ion temperature is above the twice of the electron temperature, i.e. where the ion Landau damping is large.
Fig. \ref{fig5} displays that the observed instability is of the kinetic type with $|z_{i}|\gg 1$ 
and $|z_{e}|\sim 1,5 $ for values of the $T_{e}/T_{i}$ ratio, where this instability develops. 

In Fig. \ref{fig6}, the solutions for $\omega/\omega_{pe}$, $\gamma/\omega_{pe}$, $z_{i}$ and 
$z_{e}$ versus  $\hat{U}_{ez}$ are presented. We found 
that the instability develops due to the coupled action of two effects caused by the motion 
of electrons relative to the 
practically unmovable ions in RF field. The occurrence of harmonics of the potential with 
frequencies $\omega\pm \omega_{0}$ observed in the ion 
frame is the consequence of the oscillatory motion of electrons relative to ions. The 
instability develops in the finite interval of the $\hat{U}_{ez}$ 
values and is absent in the uniform boundless plasma, where the ponderomotive current is 
absent. The growth rate maximum attains for $\hat{U}_{ez}= 2.11$ 
and is absent for $U_{ez}>4 v_{Te}$. Therefore, the instability found is the parametric 
instability driven by the ponderomotive current.

In Fig. \ref{fig7}, the solutions for $\omega/\omega_{pe}$, $\gamma/\omega_{pe}$, $z_{i}$ and 
$z_{e}$ versus the normalized effective collision frequency $
\hat{\nu}_{eff}$ are presented. The magnitude of the effective collision frequency
$\hat{\nu}_{eff}$  for a given plasma and RF field parameters should 
be derived consistently employing the nonlinear theory of the instability considered. Because the growth rate and the frequency of the 
instability are comparable, the nonlinear evolution of this instability should be studied 
using the methods of the strong plasma turbulence\cite{Galeev}.  
Figure \ref{fig7} predicts that the maximum growth rate attains 
for the comparable values of $\hat{\nu}_{eff}$ and the growth rate. Figure \ref{fig7} displays that for 
low values of $\nu_{eff}\ll 1$, i.e. high current velocity $U_{ez}$, instability is absent.

%In Fig. \ref{fig8}, the solutions versus $\kappa \lambda_{De}$ 
%are presented. It follows, that the maximum growth rate corresponds to $\kappa\lambda_{De}
%\approx 10^{-1}$. Because the maximum growth rate of the 
%instability attains, as it follows from Fig. \ref{fig3}, at $k_{z}\lambda_{De}\approx 0.7$, 
%we have $k_{z}\approx 7\kappa$ in this case.

\section{Conclusions}\label{sec5}
In this paper, the stability theory of the skin layer plasma of the inductive discharge with high RF frequency is 
developed for the case when the electron quiver velocity in the RF wave is of the order of or 
is larger than the electron thermal velocity. In this case, the electron equilibrium  distribution function (\ref{5}) 
is spatially inhomogeneous and time dependent. By employing the methodology of 
the oscillating modes, developed in Sec. \ref{sec3}, which accounts for the oscillating 
motion of the electron component relative to the  unmovable ions in the spatially inhomogeneous 
RF field of the skin layer, the governing equation (\ref{45}) for the perturbed electrostatic potential was derived. The 
numerical solution of this equation predicts the existence of the electrostatic instability of the kinetic type in the skin 
layer with the growth rate comparable with frequency. We found 
that the instability stems from the coupled action of two effects caused by the motion 
of electrons relative to ions in RF field: the occurrence of harmonics of the potential with 
frequencies $\omega\pm \omega_{0}$ observed in the ion frame and their coupling due to the ponderomotive current. 
The instability exists in the finite interval of the ponderomotive current velocity and is absent in the uniform  
boundless plasma.

%\begin{figure}[!htbp]
%\includegraphics[width=0.4\textwidth]{Fig/kappa_r}
%\caption{\label{fig8} The normalized frequency $\omega/\omega_{pe}$, normalized growth rate $
%\gamma/\omega_{pe}$, $\left|z_{i} \right|$ and $\left|z_{e} \right|$
%versus $\kappa\lambda_{De}$.}
%\end{figure}

\begin{acknowledgments}
This work was supported by National R\&D Program through the National Research Foundation of 
Korea (NRF) funded by the Ministry of Education, Science and Technology (Grant No. NRF--2018R1D1A1B07050372). 
\end{acknowledgments}

\end{document}